\documentclass[aps,prd,twocolumn,groupedaddress,showpacs,floatfix,letterpaper]{revtex4}

\usepackage{graphicx}
\usepackage{multirow}
\usepackage{amsfonts}
\usepackage{amsmath}
\usepackage{amssymb}
\usepackage{hyperref}

\begin{document}

\newcommand{\makespace}{\vspace{3 mm}}
\newcommand{\DP}{\displaystyle}

\title{Measurement of the Neutrino Neutral-Current Elastic Differential Cross Section}

\date{\today}

\author{A.~A. Aguilar-Arevalo$^{14}$, C.~E.~Anderson$^{19}$,
       A.~O.~Bazarko$^{16}$, S.~J.~Brice$^{8}$, B.~C.~Brown$^{8}$,
       L.~Bugel$^{6}$, J.~Cao$^{15}$, L.~Coney$^{6}$,
       J.~M.~Conrad$^{13}$, D.~C.~Cox$^{10}$, A.~Curioni$^{19}$,
       Z.~Djurcic$^{2}$, D.~A.~Finley$^{8}$, B.~T.~Fleming$^{19}$,
       R.~Ford$^{8}$, F.~G.~Garcia$^{8}$,
       G.~T.~Garvey$^{11}$, J.~Grange$^{9}$, C.~Green$^{8,11}$, J.~A.~Green$^{10,11}$,
       T.~L.~Hart$^{5}$, E.~Hawker$^{4,11}$,
       R.~Imlay$^{12}$, R.~A. ~Johnson$^{4}$, G.~Karagiorgi$^{13}$,
       P.~Kasper$^{8}$, T.~Katori$^{10,13}$, T.~Kobilarcik$^{8}$,
       I.~Kourbanis$^{8}$, S.~Koutsoliotas$^{3}$, E.~M.~Laird$^{16}$,
       S.~K.~Linden$^{19}$, J.~M.~Link$^{18}$, Y.~Liu$^{15}$,
       Y.~Liu$^{1}$, W.~C.~Louis$^{11}$,
       K.~B.~M.~Mahn$^{6}$, W.~Marsh$^{8}$, C.~Mauger$^{11}$,
       V.~T.~McGary$^{13}$, G.~McGregor$^{11}$,
       W.~Metcalf$^{12}$, P.~D.~Meyers$^{16}$,
       F.~Mills$^{8}$, G.~B.~Mills$^{11}$,
       J.~Monroe$^{6}$, C.~D.~Moore$^{8}$, J.~Mousseau$^{9}$, R.~H.~Nelson$^{5}$,
       P.~Nienaber$^{17}$, J.~A.~Nowak$^{12}$,
       B.~Osmanov$^{9}$, S.~Ouedraogo$^{12}$, R.~B.~Patterson$^{16}$,
       Z.~Pavlovic$^{11}$, D.~Perevalov$^{1,8}$, C.~C.~Polly$^8$, E.~Prebys$^{8}$,
       J.~L.~Raaf$^{4}$, H.~Ray$^{9,11}$, B.~P.~Roe$^{15}$,
       A.~D.~Russell$^{8}$, V.~Sandberg$^{11}$, R.~Schirato$^{11}$,
       D.~Schmitz$^{6}$, M.~H.~Shaevitz$^{6}$, F.~C.~Shoemaker$^{16}$\footnote{deceased},
       D.~Smith$^{7}$, M.~Soderberg$^{19}$,
       M.~Sorel$^{6}$\footnote{Present address: IFIC, Universidad de Valencia and CSIC, Valencia 46071, Spain},
       P.~Spentzouris$^{8}$, J.~Spitz$^{19}$, I.~Stancu$^{1}$,
       R.~J.~Stefanski$^{8}$, M.~Sung$^{12}$, H.~A.~Tanaka$^{16}$,
       R.~Tayloe$^{10}$, M.~Tzanov$^{5}$,
       R.~G.~Van~de~Water$^{11}$,
       M.~O.~Wascko$^{12}$\footnote{Present address: Imperial College; London SW7 2AZ, United Kingdom},
        D.~H.~White$^{11}$,
       M.~J.~Wilking$^{5}$, H.~J.~Yang$^{15}$,
       G.~P.~Zeller$^8$, E.~D.~Zimmerman$^{5}$ \\
\smallskip
(The MiniBooNE Collaboration)
\smallskip
}
\smallskip
\smallskip
\affiliation{
$^1$University of Alabama; Tuscaloosa, AL 35487 \\
$^2$Argonne National Laboratory; Argonne, IL 60439 \\
$^3$Bucknell University; Lewisburg, PA 17837 \\
$^4$University of Cincinnati; Cincinnati, OH 45221\\
$^5$University of Colorado; Boulder, CO 80309 \\
$^6$Columbia University; New York, NY 10027 \\
$^7$Embry Riddle Aeronautical University; Prescott, AZ 86301 \\
$^8$Fermi National Accelerator Laboratory; Batavia, IL 60510 \\
$^9$University of Florida; Gainesville, FL 32611 \\
$^{10}$Indiana University; Bloomington, IN 47405 \\
$^{11}$Los Alamos National Laboratory; Los Alamos, NM 87545 \\
$^{12}$Louisiana State University; Baton Rouge, LA 70803 \\
$^{13}$Massachusetts Institute of Technology; Cambridge, MA 02139 \\
$^{14}$Instituto de Ciencias Nucleares, Universidad National Aut\'onoma de M\'exico, D.F. 04510, M\'exico \\
$^{15}$University of Michigan; Ann Arbor, MI 48109 \\
$^{16}$Princeton University; Princeton, NJ 08544 \\
$^{17}$Saint Mary's University of Minnesota; Winona, MN 55987 \\
$^{18}$Virginia Polytechnic Institute \& State University; Blacksburg, VA 24061 \\
$^{19}$Yale University; New Haven, CT 06520\\
}

\begin{abstract}
We report a measurement of the flux-averaged neutral-current elastic
differential cross section for neutrinos scattering on mineral oil (CH$_2$) as a function of four-momentum transferred squared.
It is obtained by measuring the kinematics of recoiling nucleons with kinetic energy greater than 50~MeV which are readily detected in MiniBooNE.
This differential cross-section distribution is fit with fixed nucleon form factors apart from an axial mass, $M_A$, 
that provides a best fit for $M_A= 1.39\pm0.11$~GeV.
Additionally, single protons with kinetic energies above 350 MeV can be distinguished from neutrons and multiple nucleon events.
Using this marker, the strange quark contribution to the neutral-current axial vector form factor at $Q^2 = 0$, $\Delta s$, 
is found to be $\Delta s=0.08\pm0.26$.
\vspace{1.5in}
\end{abstract}


\maketitle

\pagebreak[4]


\section{Introduction.}



Neutrino-nucleon neutral-current elastic (NCE) scattering is a unique
and fundamental probe of the nucleon. NCE scattering on a nuclear
target such as carbon may be viewed as scattering from the individual
nucleons but may also include contributions from collective nuclear
effects. This process should be sensitive to nucleon isoscalar weak
currents as opposed to charged-current quasi-elastic (CCQE) scattering
which interacts only via isovector weak currents. Therefore, the NCE
process can be used to search for strange quarks in the nucleon which
may show themselves via the isoscalar weak current. In addition, the
NCE process offers a complementary channel to CCQE to investigate any
substantial collective nuclear effects in a nucleus such as carbon.

The MiniBooNE experiment~\cite{MB_beam, MB_detec} searches for neutrino oscillations at
Fermilab and NCE events account for 18\% of the total neutrino sample
collected. A large fraction of these neutral-current events are
readily observable in the MiniBooNE detector which uses pure mineral
oil (CH$_2$) as a detector medium. The detector is predominantly a Cherenkov
detector, however, fluors presented in the mineral oil produce
a small amount of scintillation light well below Cherenkov threshold
for relativistic charged particles (such as electrons, muons, protons,
etc.). The absence of prompt Cherenkov light allows for the
identification and measurement of the recoiling nucleons produced in
neutrino NCE scattering.

The main result presented in this paper is a high-statistics measurement of the
flux-averaged differential cross section as a function of $Q^2$ for NCE scattering 
on CH$_2$ in MiniBooNE for $0.1$~GeV$^2<Q^2<1.65$~GeV$^2$.
It is presented as scattering from individual nucleons
both bound (in carbon) and free (in hydrogen). However, it is acknowledged
that nuclear effects may well be important for full understanding of this data. 

A preliminary measurement of the MiniBooNE NCE differential cross section had been performed
early on using $6.57\times10^{19}$ protons on the neutrino production target (POT)~\cite{ChrisCox}.
The results presented in this paper are based on the entire data set corresponding in neutrino run mode, 
for a total of $6.46\times10^{20}$ POT and a different event reconstruction.
MiniBooNE has also collected anti-neutrino data which will be the presented in subsequent works.

To characterize the NCE from carbon in MiniBooNE, the relativistic Fermi gas (RFG) 
model of Smith and Moniz is used as described in Section~\ref{sec:xs_model}.
Within this model, two fundamental parameters are employed: the nucleon 
axial mass, $M_A$, and the strange quark contribution to the axial form
factor, $\Delta s$. It is acknowledged that nuclear effects in carbon may 
cause this axial mass to be an ``effective'' value and not the same as that
for scattering from free nucleons. The formalism for NCE scattering 
on free nucleons, which is basic to any model-dependent approach to 
characterize the NCE data, is summarized in Appendix~B.

While the generally accepted value for $M_A$ is $1.026 \pm 0.021$
GeV~\cite{MA_ave}, recent experiments ~\cite{CCQE_PRL,K2K_ccqe,
K2K_ccqe2, MinosMA_Dorman_Thesis} measuring CCQE from nuclear targets 
have found it useful to employ values that are 20--30\% larger to fit the 
$Q^2$ (squared four-momentum transfer) dependence of their
observed yields. It may be that this increased value of $M_A$ 
should be understood not as the $M_A$ obtained for free nucleons 
but rather as a parameterization of neglected nuclear effects~\cite{Martini_CCQE,Benhar_CCQE}.
Regardless, an extraction of $M_A$ from NCE scattering offers a complementary
test to the $M_A$ determined from CCQE.

The strange quark contribution to the nucleon spin at $Q^2=0$, $\Delta s$, can be extracted 
via the NCE process within an RFG model. The NCE differential 
cross section at low $Q^2$ is sensitive to $\Delta s$~\cite{Garvey1} for both
neutrinos and antineutrinos. The BNL E734 experiment measured these
processes and reported $\Delta s$~\cite{BNL734,Garvey_Louis,BNLReanalysis_Alberico}.
However, ratio measurements offer the possibility for extracting $\Delta s$ 
at low $Q^2$ with reduced systematic errors. For example, a measurement of 
$(\nu p\to\nu p)/(\nu_\mu n\to\mu p)$, has been proposed by the FINeSSE experiment~\cite{FINeSSE_proposal}.
In this paper, a ratio of $(\nu p\to\nu p)/(\nu N\to\nu N)$ at $Q^2 > 0.7$~GeV is used 
to extract $\Delta s$, where $N$ is a neutron or proton.

In the following sections the MiniBooNE experiment is described,
including a description of the neutrino beamline, detector, neutrino flux prediction, 
and the cross-section model used to predict the rates of different neutrino interactions in the detector.
In Sections~\ref{sec:nce_analysis} and~\ref{sec:deltas_measurement}, methods and techniques used 
in the NCE analysis are presented together with various results,
including the NCE differential cross-section, the NCE/CCQE ratio and measurements of $M_A$ and $\Delta s$.
Section~\ref{sec:summary} contains a summary of the paper.
Also, Appendices A and B have additional information 
useful for interpreting the results which is not included in the main text.

%
\section{MiniBooNE experiment}\label{sec:mb_experiment}
\subsection{Neutrino Beamline.}
The Booster Neutrino Beamline (BNB) at the Fermi National Accelerator
Laboratory uses a beam of protons with momentum $8.89\mbox{ GeV}/c$ to produce 
an intense and almost pure beam of $\nu_\mu$ with an average energy of about 800~MeV.
Protons are extracted from the Fermilab Booster in 1.6~$\mu$s pulses with
$\sim 4\times10^{12}$ protons in each beam pulse.
They are delivered onto a beryllium target, where a secondary beam of mesons is
produced in p-Be interactions.
Mesons are passed through a magnetic horn, a device which focuses positively charged particles and defocuses negatively charged particles.
Mesons decay in an air-filled decay pipe producing a beam of neutrinos. 
Using the magnetic horn increases the neutrino flux at the MiniBooNE detector by a factor of $\sim6$.
The details on the BNB components can be found in Ref.~\cite{MB_beam}.

\subsection{Neutrino Flux Prediction.}\label{sec:flux}
The neutrino flux at the detector is calculated via a GEANT4-based~\cite{Geant4} MC beam simulation.
The simulation includes a full beam geometry, specified by shape, location and
material composition of the BNB components.
The MC generates protons upstream of the target and propagates them through the
target, generating and propagating products of p-Be interactions through the
rest of the simulated BNB.
In the $\nu_\mu$ flux at the MiniBooNE detector, 96.7\% of neutrinos are produced via $\pi^+\to\mu^+ + \nu_\mu$ decay.
The $\pi^+$ production double differential cross section used in the beam MC is
based on a fit to an external measurement from the HARP experiment on the
same target and with the same proton beam energy as in the BNB~\cite{HARP}.

\begin{figure}[t!]
\includegraphics[width=\columnwidth]{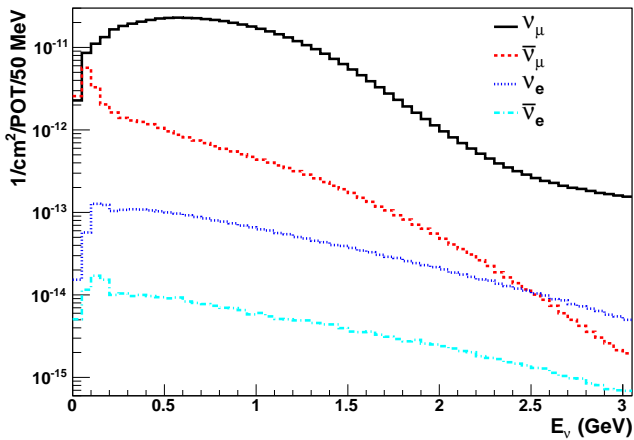}
\vspace{-2mm}
\caption{Neutrino flux at the MiniBooNE detector for different types of
         neutrinos as a function of their energy as reported in~\cite{MB_beam} and~\cite{MB_opendata_flux}. }
\label{fig:nutotflux}
\end{figure}

The neutrino flux prediction for different types of neutrino species is shown
in Fig.~\ref{fig:nutotflux}.
Flux tables are available in Ref.~\cite{MB_opendata_flux}.
The $\pi^{+}$ production contribution to the neutrino flux uncertainty is
about 5\% at the peak of the flux distribution, 
increasing significantly at low and high neutrino energies.
Other contributions to the flux error include uncertainties on other mesons
production cross sections, the number of POT, and the horn magnetic field~\cite{MB_beam},~\cite{MB_CCQE_PRD}.

\subsection{Detector.}
The MiniBooNE detector is situated 541~m from the Be target, under 3~m of
overburden in order to reduce cosmic backgrounds.
It is a 12.2~m diameter spherical steel tank, filled with mineral oil.
The tank is divided into two optically isolated regions, a \textit{signal} region being an inner sphere of radius 5.75 m, 
and a \textit{veto} region that is an outer shell with a thickness of 0.35 m.
Photomultiplier tubes (PMTs) are used to detect photons emitted 
by the charged particles which are produced in the neutrino interactions.
Charged particles may emit both Cherenkov and scintillation light.
Information from all PMTs is used to identify and reconstruct the products of neutrino interactions.

A total of 1520 8-inch PMTs~\cite{PMT} are instrumented in the detector.
There are 1280 PMTs attached to the spherical barrier from the inside, in the
signal region, facing toward the center of the tank and distributed
approximately uniformly.
The remaining 240 PMTs are placed in the veto region, and are used
to tag charged particles entering or leaving the tank.
The veto PMTs are mounted back-to-back, tangentially to the optical barrier, in
order to have as much veto view as possible.
Details on the MiniBooNE detector can be found elsewhere~\cite{MB_detec}.

\subsection{Cross-Section Model.}\label{sec:xs_model}
Neutrino interactions within the detector are simulated with the NUANCE-v3
event generator~\cite{nuance}, where the relativistic Fermi gas model of Smith
and Moniz~\cite{SmithMoniz} is used to describe NCE scattering.
Fermi momentum for carbon is taken to be $220\pm20$~MeV and binding energy $34\pm9$~MeV.

The contribution from strange quarks to the vector and axial vector form factors is taken to be zero.
The error on $\Delta s$ id taken to be 0.1.
The value of $M_A$ used in the MC is different for the quasi-elastic (both
neutral and charged current) scattering on carbon and hydrogen: for scattering
on carbon $M_A=1.230\pm0.077$~GeV is used (as measured from the CCQE channel in
MiniBooNE~\cite{CCQE_PRL}), while for scattering on hydrogen $M_A=1.13\pm0.10$~GeV is used (which is
the average between the values measured by the deuterium-based scattering experiments and MiniBooNE).

For resonant pion production, the Rein and Sehgal model~\cite{ReinSehgal} is used. 
In the few GeV range, such processes are dominated by the $\Delta(1232)$ resonance,
although contributions from higher mass resonances are also included in the MC.
The values $M_A^{1\pi} = 1.10\pm0.27$~GeV for both charge current (CC) and NC single pion events, 
and $M_A^{N\pi} = 1.30\pm0.52$~GeV for both CC and NC multiple pion events are used.

Intranuclear final state interactions (FSI) 
inside the carbon nucleus are modeled in NUANCE using a binary cascade model~\cite{nuance},
where the scattered hadrons are propagated through the nucleus, which is
simulated based on models of nuclear density and Fermi momentum.
Due to FSI, a NCE interaction may produce more than one final state particle (other than the neutrino).
For NCE scattering on carbon, the probability of producing multiple nucleons is $\sim26\%$ integrated
over the MiniBooNE flux, according to NUANCE.
Also, a NC pion event might not contain any pions in the final state as the pion
can be absorbed in the carbon nucleus or the baryonic resonance re-interacts without decaying.
These are the dominant mechanisms by which NC pion events can become backgrounds
to this analysis. The probability that a pion is absorbed is $\sim20\%$ in carbon
for MiniBooNE energies, according to the NUANCE simulation.
The intranuclear pion absorption cross section is assigned a 25\% uncertainty based
on external pion-carbon data~\cite{piabs1,piabs2,piabs3},
and $\Delta N\to NN$ interactions are assigned a 100\% uncertainty.

Neutrino interactions outside the detector, in the surrounding dirt or in
the detector material (referred to as ``dirt'' background henceforth), are
simulated the same way as the in-tank interactions but with a cross section
reweighted according to the density of the material relative to that of the mineral oil.

Particle propagation in the detector is modeled using a GEANT3-based~\cite{GEANT3} MC with
GCALOR~\cite{gcalor} hadronic interactions simulating the detector response to
particles produced in the neutrino interactions.
GCALOR was chosen over GFLUKA~\cite{GFLUKA} simulation, 
as it provides a better model of $\pi^+$ interactions on carbon~\cite{Kendall_thesis}.

%
\section{Neutral--Current Elastic Analysis.}\label{sec:nce_analysis}
\subsection{Event Reconstruction.}
Event reconstruction in MiniBooNE is based on finding a set of parameters 
(position, time, direction and energy -- where applicable)
which maximizes the event likelihood using the charge and time information from all PMTs.
Each event is reconstructed using some combination of six different event hypotheses
-- single proton (NCE-like), single muon ($\nu_\mu$ CCQE-like), single electron ($\nu_e$ CCQE-like), 
single $\pi^0$ (NC $\pi^0$ production-like), muon and $\pi^+$ with the same vertex 
($\nu_\mu$ CC $\pi^+$ production-like), and
muon and $\pi^0$ with the same vertex ($\nu_\mu$ CC $\pi^0$ production-like).
A charge-time likelihood minimization method~\cite{Perevalov_thesis} is used to obtain
the best estimate of the kinematic observables in each event hypothesis.
Under the NCE hypothesis, each event is assumed to be a point-like proton 
with Cherenkov and scintillation light emission profiles determined from the MC.
The output variables from these event reconstructions
(such as likelihood ratios between two different event hypotheses) allow for particle identification.

The resulting position resolution is~$\sim0.75$~m for proton events in the detector and
$\sim1.35$~m for neutrons, with an energy resolution of~$\sim 20\%$ for protons 
and~$\sim 30\%$ for neutrons.
For protons above Cherenkov threshold, the direction resolution is~$\sim 10^{\circ}$.

\begin{figure}
\includegraphics[bb = 0 0 568 385, width=\columnwidth]{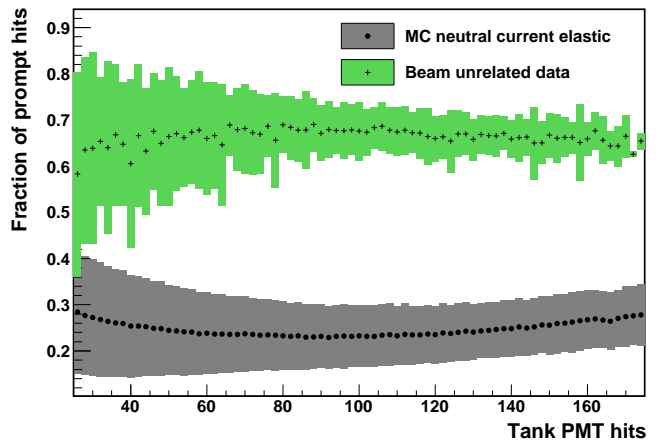}
\caption{Fraction of prompt hits versus the total number of tank PMT hits for
         beam unrelated data and NCE MC events reconstructed under an electron
         hypothesis.
         The error bars correspond to the RMS of the distributions.}
\label{fig:elfra}
\end{figure}

Light emission properties of protons differ from those of other charged particles in the detector, 
allowing for their particle identification. 
For instance, protons differ from electrons in terms of the fraction of prompt light emitted, 
defined as the fraction of PMT hits with
a corrected time between $-5$~ns and 5~ns, as illustrated in Fig.~\ref{fig:elfra},
where the corrected time is the time difference between the PMT hit time and 
the reconstructed event time, with light propagation time from the reconstructed vertex to the PMT also taken into account.

Being neutral particles, neutrons themselves do not cause light emission in the detector.
However we may detect them through their subsequent strong re-interactions, 
in which usually energetic protons are produced.
Because we detect NCE neutrons only through secondary protons, they are virtually
indistinguishable from NCE proton events.

For MiniBooNE NCE interactions, the total charge on all PMTs is proportional to
the sum of kinetic energies of all final state nucleons that are produced in the interaction, 
which is referred to throughout this paper as $T$.
It is important to understand that the nucleon kinetic energy measured this way is different from the one determined from
the track-based reconstruction used in the SciBooNE~\cite{Takei_Phd} and BNL E734~\cite{BNL734} experiments. 
In that case, the reconstructed proton track length is proportional to the kinetic energy of the most energetic proton produced in the event.
Also the particle identification in MiniBooNE is based almost entirely on the properties
of the measured Cherenkov ring (such as ring sharpness, charge and time likelihoods), 
whereas the track-based experiments mostly use the particle's energy loss along the track.

%
\subsection{Event Selection.}\label{sec:cuts}
\begin{figure}
\includegraphics[bb = 5 0 560 385, width=\columnwidth]{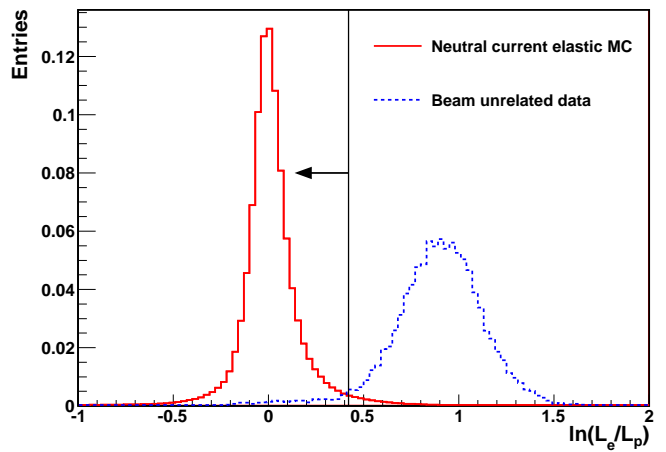}
\caption{Log-likelihood ratio between electron and proton event hypotheses for
         MC-generated NCE scattering events and beam unrelated data.
         Both histograms are normalized to unit area.
         Events with $\ln(\mathcal{L}_e/\mathcal{L}_p)<0.42$ are selected for the analysis.}
\label{fig:tllkdiff}
\end{figure}

The following set of selection criteria (cuts) are applied to the full MiniBooNE data set
to select the NCE sample:
\begin{enumerate}
\item Only 1 $subevent$ to ensure the event is NC and includes no decaying particles (e.g., $\mu$ decay).
      A $subevent$ is a cluster of at least 10 tank PMT hits for which there is
      no more than 10~ns between any two consecutive hits.
      For example, a CCQE event typically contains two subevents;
      the first subevent is associated with the outgoing muon,
      while the second is associated with the subsequent decay electron~\cite{MB_CCQE_PRD}.
\item Number of veto PMT hits is less than 6 to remove events exiting
      or entering the detector, $VHits<6$. 
\item Number of tank PMT hits is greater than 24 to ensure a reliable reconstruction, $THits>24$.
\item Beam time window cut in order to consider only events time-coincident with the neutrino beam.
\item Reconstructed proton energy of $T<650$~MeV (above which the
      signal to background ratio decreases significantly).
\item Log-likelihood ratio between electron and proton event hypotheses of $\ln(\mathcal{L}_e/\mathcal{L}_p)<0.42$.
      The purpose of this cut is to eliminate beam-unrelated electrons from cosmic-ray muon decays.
      Even though the variable shown in Fig.~\ref{fig:elfra} provides a good separation between proton and electron events,
      it was found that the log-likelihood ratio is most effective in capturing all
      differences between the NCE signal and this background.
      The $\ln(\mathcal{L}_e/\mathcal{L}_p)$ variable and the value of the cut are shown in Fig.~\ref{fig:tllkdiff}
      for simulated NCE and the beam unrelated data.
      The beam unrelated data events are mostly muon decay (Michel) electrons.
\item A fiducial volume cut, defined as follows:
      $$ R_{fiducial} (T) = \begin{cases} 
                R<4.2\,\mbox{m}  & \mbox{if}\quad T<200\,\mbox{MeV},\\
                R<5.0\,\mbox{m}  & \mbox{if}\quad T>200\,\mbox{MeV}.
                \end{cases} $$
          A tighter fiducial volume is required at low energies to reduce the dirt background.
\end{enumerate}

\begin{figure}
\includegraphics[bb = 0 0 568 385, width=\columnwidth]{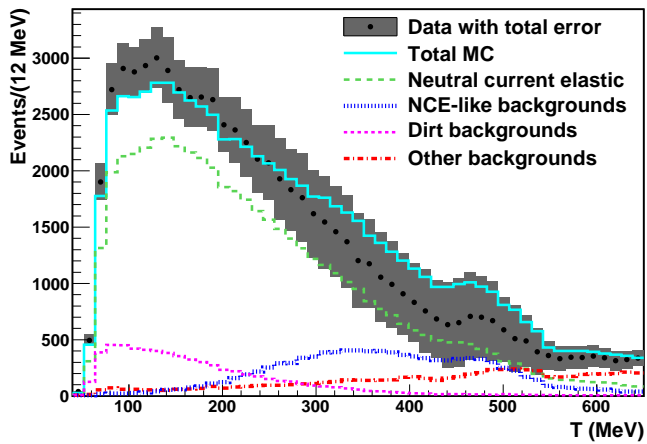}
\caption{Reconstructed nucleon kinetic energy spectra for the data and MC
         after the NCE event selection and a uniform fiducial volume cut of $R<4.2$~m are applied.
         All MC distributions are normalized to the number of protons on target (POT).}
\label{fig:erec}
\end{figure}

A total of 94,531 events pass the NCE cuts resulting from $6.46\times10^{20}$ POT.
This is the largest NCE event sample collected to date.
The efficiency of the cuts is estimated to be 35\%;
a large portion of which stems from the fiducial volume cut.
We consider all NCE events with original vertices inside the detector as \textit{signal}.
The predicted fraction of NCE events in the sample is 65\%.
The remaining 35\% of events are \textit{backgrounds} of different types:
15\% are NCE-like backgrounds,
10\% dirt events,
and 10\% other backgrounds (of which only 0.5\% are beam unrelated).
The reconstructed nucleon kinetic energy spectrum for selected NCE events
with a uniform fiducial volume cut is shown in Fig.~\ref{fig:erec} along with the predicted background
contributions.

The NCE-like background consists of NC pion production channels with no pion in
the final state (i.e. the pion is absorbed in the initial target nucleus through FSI).
In this case, the final state particles for these events are solely nucleons.
In MiniBooNE, this is indistinguishable from the final state produced in NCE
events, and hence why these events are referred to as NCE-like background.
The NCE-like background contributes mostly at intermediate energies, $200\mbox{ MeV}<T<500$ MeV.

The dirt background is an important contribution to the NCE data sample at low
energies, most significantly below 200~MeV.
This background is due to nucleons (mainly neutrons) which are
produced in neutrino interactions outside of the detector, penetrating into the
detector without firing enough veto PMTs.
Dirt events are challenging to simulate in the MC because they occur in various
media that have not been studied in detail (in the soil, detector support
structures, etc.).
However, we directly constrain this background using MiniBooNE data.
Dirt events can be isolated from in-tank interactions using their distinct
kinematics: dirt events are preferentially reconstructed in the most upstream
($Z<0$~m) and outer regions of the detector with relatively low energies (small
values of $T$).
The dirt energy spectrum is measured by fitting dirt-enriched samples in the
variables $Z$, $R$ and $T$, as explained in detail in Appendix~A.

Other backgrounds are mainly charged-current channels, but also include neutral
current pion production, beam unrelated, and anti-neutrino NCE events.
These backgrounds become relevant at high reconstructed energies, mainly above
400~MeV.

\subsection{Unfolding.}
After subtracting the beam unrelated and dirt backgrounds,
the reconstructed energy spectrum for the data is multiplied by the signal
fraction, which is the number of NCE events divided by the total number of neutrino induced
in-tank events according to MC prediction, bin by bin.
The obtained NCE reconstructed energy spectrum is then corrected for detector resolution and
efficiency effects that distort the original spectrum.
The correction is applied by unfolding the distribution using a method based on Bayes' theorem~\cite{Agostini}. 

The unfolding matrix applied to the reconstructed spectrum is calculated for
NCE events in the MC using a migration matrix of the true nucleon kinetic
energy (the sum of the kinetic energies of all nucleons in the final state)
versus the reconstructed nucleon kinetic energy.
This method gives a well-behaved but biased solution, which depends on the
original MC energy spectrum.
The error due to the unfolding bias is estimated by iterating the unfolding procedure
for each MC variation, 
where the new MC energy spectrum is replaced by the unfolded energy spectrum.
The details of the unfolding procedure used in the measurement and the error estimation associated with its bias can be found in Ref.\cite{Perevalov_thesis}.
The detector resolution and cut efficiency effects can be seen by comparing the reconstructed energy spectrum before unfolding (Fig.~\ref{fig:erec}) 
and the final cross-section result (Fig.~\ref{fig:xs}), which we shall discuss later.

\subsection{Neutral--Current Elastic Flux-Averaged Cross Section.}\label{sec:xs}

For each NCE event, $Q^2$ can be determined by measuring the total kinetic energy 
of outgoing nucleons in the interaction assuming the target nucleon is at rest.
In this case we define:
$$Q^2_{QE} = 2m_N T = 2m_N \sum_i T_i,$$  
where $m_N$ is the nucleon mass and $T$ is the sum of the kinetic energies of the final state nucleons.
MiniBooNE reports the NCE differential cross section as a function of of this variable, $Q^2_{QE}$.

The MiniBooNE NCE differential cross-section is less sensitive to FSI effects than those measured by tracking detectors, such as BNL E734~\cite{BNL734}. 
In case of a FSI, where no pions in the final state have been produced, the energy transferred by the neutrino may be divided among several outgoing nucleons, 
but the total energy released in the MiniBooNE detector stays roughly the same due to energy conservation.
Track-based detectors measure $Q^2$ by the proton track length and its angle with respect to the beam direction, 
which are kinematic observables of the most energetic proton produced in the NCE event.
In that case, FSI may have large effects on the kinematics of individual outgoing nucleons, including the most energetic nucleon.
Of course, there are still some FSI interactions producing final state pions which must be modelled and which can affect MiniBooNE NCE cross-section measurement.

\begin{figure*}
\includegraphics[bb = 0 0 565 385, width=1.5\columnwidth]{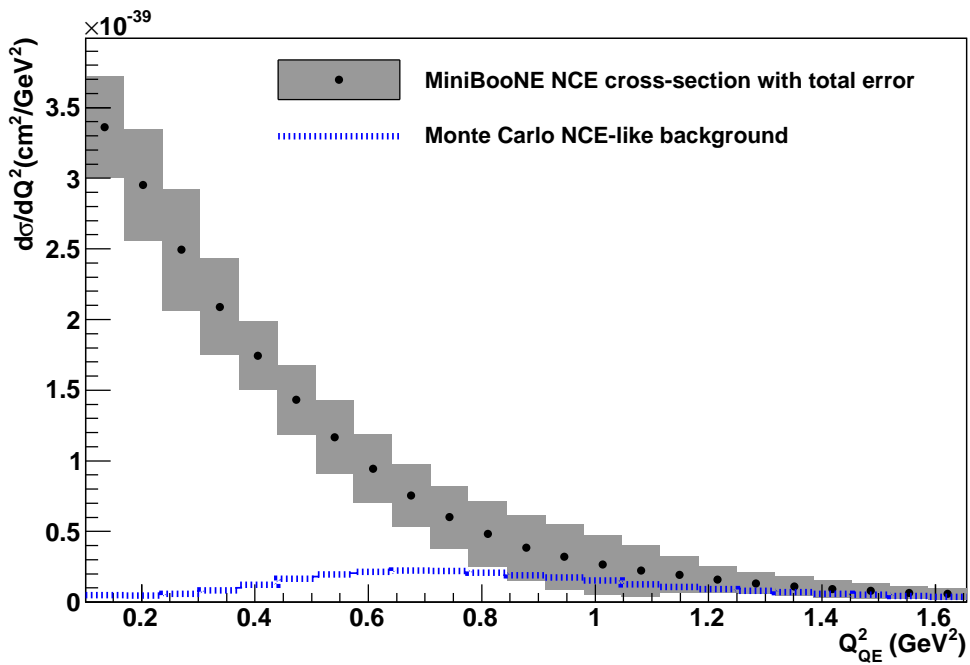}
\caption{ The MiniBooNE NCE ($\nu N \to \nu N$) flux-averaged differential cross section on CH$_2$ as a function of $Q^2_{QE} = 2m_N\sum_i T_i$, where we sum the true kinetic energies of all final state nucleons produced in the NCE interaction.
The blue dotted line is the predicted spectrum of NCE-like background which has been subtracted out from the total NCE-like differential cross section.}
\label{fig:xs}
\end{figure*}

The resulting NCE flux-averaged differential cross section on CH$_2$ is shown in
Fig.~\ref{fig:xs} as a function of $Q^2_{QE}$.
The predicted distribution of the NCE-like background, which has been
subtracted along with the rest of backgrounds, is also shown in the figure.
The NCE scattering is a sum of three different processes: scattering on
free protons in hydrogen, bound protons in carbon, and bound neutrons in carbon.
A detailed description of the contributions of each of these processes to the 
total MiniBooNE NCE cross section is given in Appendix~B.

Systematic uncertainties and their contribution to the total error have been studied. 
The normalization error can be represented by a single number, shown in Table~\ref{tab:errors}.
The largest systematic error in the NC analysis, the optical model, arises from the uncertainty
on both Cherenkov and scintillation light production by charged particles in mineral oil.
Cherenkov light production has been calibrated on cosmic muons, Michel electrons and neutral pions~\cite{MB_detec}.
An error of 20\% on this amplitude of scintillation light generated by sub-Cherenkov particles in the detector has been assigned
through a combination of benchtop measurements and {\it{in situ}} calibrations (see Appendix C of Ref~\cite{Perevalov_thesis} for details).

\begin{table}
\begin{tabular}{l|l|r}
\hline
\hline
Type of error&    Error  & Value (\%)\\
\hline
 Flux & Flux                                                 & 6.7  \\
\hline
\multirow{3}{*}{Cross section}  & NC $\pi^0$ production yield   & 0.5         \\
  & In-tank background cross section   & 6.3         \\
  & Dirt background                                  & 1.0  \\
\hline
\multirow{4}{*}{Detector}  &  Discriminator threshold & 0.6  \\
  &  Optical model                               & 15.4 \\
  &  Charge-time PMT response     & 2.1  \\
  & Hadronic interactions                & 0.5  \\
\hline
Total  &                                              & 18.1 \\
\hline
\hline
\end{tabular}
\caption{Individual error contributions to the total integrated normalization uncertainty on the MiniBooNE measured NCE cross section.
The statistical error of 2.5\% is not included in the total normalization error.
}
\label{tab:errors}
\end{table}

\subsection{Neutral--Current Elastic to Charged--Current Quasi-Elastic Cross-Section Ratio Measurement.}\label{sec:nce_ccqe_ratio}

Given that MiniBooNE measures a CCQE differential cross section~\cite{MB_CCQE_PRD} that is $\sim30\%$ higher 
than naive expectations from the relativistic Fermi gas model~\cite{SmithMoniz}, it is
interesting to compare those results with our NCE measurement.
To facilitate such a comparison and, at the same time, reduce flux uncertainties, we extract the NCE/CCQE ratio as a function of $Q^2_{QE}$.
In the case of CCQE, $Q^2_{QE}$ has been defined from the outgoing muon kinematics only, 
assuming a stationary neutron target (see Ref.~\cite{MB_CCQE_PRD} for details).
It should be pointed out that a significant difference exists in how these cross sections are measured in MiniBooNE.
As explained earlier, the NCE cross section is calculated from the measured total kinetic energy of final state nucleons and is mildly sensitive to FSI.
Whereas the CCQE is calculated entirely from the reconstructed muon and is not sensitive to FSI.

\begin{figure}[t!]
\includegraphics[width=\columnwidth]{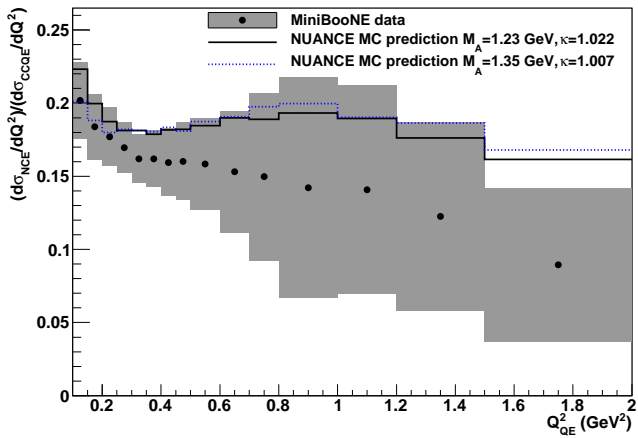}
\vspace{-2mm}
\caption{ MiniBooNE NCE/CCQE cross section ratio on CH$_2$ as a function of $Q^2_{QE}$.
The NUANCE MC prediction is plotted for two assumptions of parameters: 
black solid line for $M_A = 1.23$~GeV and $\kappa=1.022$~\cite{CCQE_PRL}, and blue dotted line for $M_A = 1.35$~GeV and $\kappa=1.007$~\cite{MB_CCQE_PRD}.
$\Delta s$ is assumed to be zero in both of these cases.
Both cross sections in the ratio are per target nucleon -- there are 14/6 times more target nucleons in the numerator than in the denominator.
The error bars represent both statistical and all systematic uncertainties (excluding flux errors) taken in quadrature. }
\label{fig:nce_ccqe_ratio}
\end{figure}

The measured ratio is shown in Fig.~\ref{fig:nce_ccqe_ratio} together with the NUANCE MC prediction. 
The data/MC agreement is reasonable within errors.

\begin{figure}[t!]
\includegraphics[width=\columnwidth]{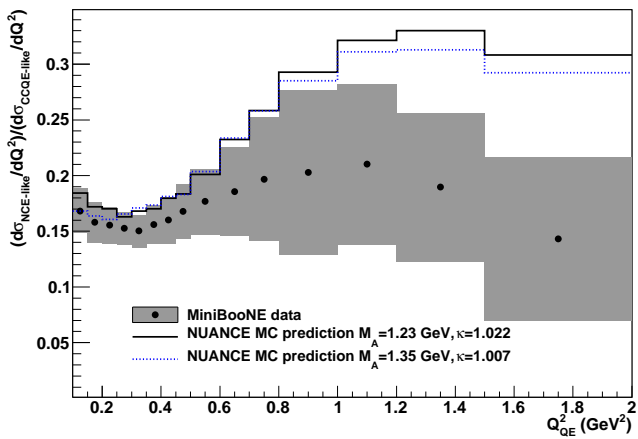}
\vspace{-2mm}
\caption{MiniBooNE NCE--like/CCQE--like cross-section ratio on CH$_2$ as a function of $Q^2_{QE}$ with total error.
The color code for the histograms is the same as in the Figure~\ref{fig:nce_ccqe_ratio}.
}
\label{fig:nce_ccqe_like_ratio}
\end{figure}

Adding the MC NCE-like background prediction to the numerator and the MC CCQE-like background
prediction to the denominator produces a NCE--like to CCQE--like differential cross-section ratio,
which is additionally shown in Fig.\ref{fig:nce_ccqe_like_ratio}.
This is an even more model independent measurement, where we do not have to rely on modeling of
both NCE--like and CCQE--like backgrounds and claim them as a part of the signal.

The measured NCE/CCQE ratio is consistent with that predicted by the MC.
This is an important point when considering possible explanations of the larger
than predicted value of the CCQE cross section.
The predicted MC ratio is chosen for two values of $M_A$ and $\kappa$: one is
with $M_A = 1.23$ GeV and $\kappa=1.022$ as measured in~\cite{CCQE_PRL},
the second with $M_A = 1.35$ GeV and $\kappa = 1.007$ is a from a more recent MiniBooNE CCQE result~\cite{MB_CCQE_PRD},
where $\kappa$ is a Pauli blocking scaling factor parameter.

There is some disagreement between data and MC for the NCE-like/CCQE-like ratio above $Q^2_{QE}>1.0\mbox{ GeV}^2$, 
but this is where the NCE-like backgrounds 
(predominantly NC pion channels with pion absorption) become a significant fraction of the signal.

\subsection{Axial Vector Mass Measurement Using the NCE Cross Section.}
In the Appendix~B.1 the expression for NCE differential cross section on free nucleons is described.
From there, one can see that the NCE cross section is sensitive to the axial form factor.
In fact, at low $Q^2$, $d\sigma /dQ^2\sim (F^Z_A)^2(Q^2)$,
where
$$
F_A^Z(Q^2) = \frac12(g_A\tau_3 - \Delta s)/\left(1+Q^2/M_A^2\right)^2,
$$
where $g_A=1.2671$ is measured precisely from neutron beta decay~\cite{PDG}.

To first order, the MiniBooNE $\nu N\to \nu N$ cross section is not
sensitive to $\Delta s$, as the linear term in $\Delta s$ nearly cancels, while
the quadratic term in $\Delta s$ remains, but is small if $|\Delta s|\ll g_A$.
However, these data are still useful for probing $M_A$.

\begin{figure}
\includegraphics[bb = 0 0 565 385, width=\columnwidth]{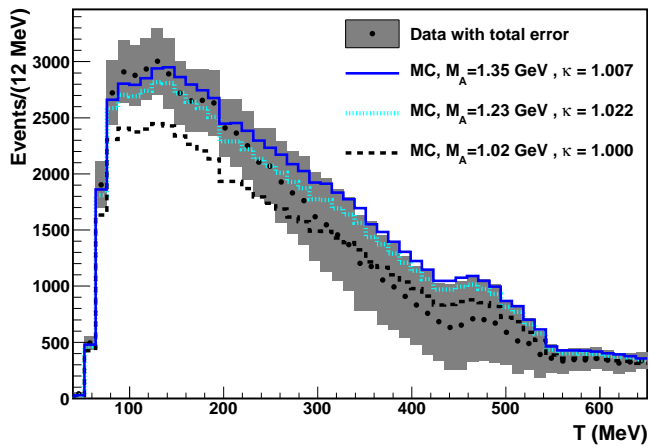}
\caption{$\chi^2$-tests performed on the NCE reconstructed nucleon kinetic energy distribution for MC with different $M_A$ values of
         1.35, 1.23, and 1.02~GeV. 
         The $\chi^2$ values are 27.1, 29.2 and 41.3 for 49 degrees of freedom (DOF), respectively.
         The distributions are normalized to POT.}
\label{fig:MA_results}
\end{figure}

A $\chi^2$ goodness of fit test is performed to find the set of $M_A$ and $\kappa$
parameters, described earlier in Section~B.2, that best matches data.
Varying values of $M_A$ and $\kappa$ in the MC model results in different
reconstructed NCE energy distributions.
For each set of $M_A$ and $\kappa$ values, a $\chi^2$ is calculated using the full error matrix.
The full error matrix in this case also includes an additional uncertainty on $\Delta s$ of 0.1.
The reconstructed energy spectrum for data and MC for different values of $M_A$
and $\kappa$ (as measured from MiniBooNE CCQE data, as well as the average
values prior to MiniBooNE) are shown in Fig.~\ref{fig:MA_results}.
The reconstructed energy distribution of the NCE sample has a negligible sensitivity to $\kappa$;
however, the higher values of $M_A$ seem to better describe MiniBooNE NCE data.

The NCE data can also be directly fit to independently extract information on $M_A$ and $\Delta s$.
Because of the quadratic term in $\Delta s$ in the NCE differential cross-section expression, the
shape of $\chi^2$ slightly depends on the value of $\Delta s$.
Assuming $\Delta s=0$, the $1\sigma$ allowed region of $M_A$ from the
MiniBooNE NCE sample yields:

$$ M_A=1.39\pm0.11\mbox{ GeV}, $$
with $\chi^2_{min}/DOF=26.9/50$.
Using $\Delta s=-0.2$ (which roughly corresponds to the value obtained by the
BNL E734 experiment~\cite{BNL734}) yields $M_A=1.35\pm0.11$~GeV with
$\chi^2_{min}/DOF=24.9/50$.
The results from the $M_A$ fit to the NCE data using an absolute (POT) normalization
agree well with the shape-only fit results from the MiniBooNE CCQE data~\cite{MB_CCQE_PRD}.

\section{Measurement of $\Delta s$ Using a High Energy Proton-Enriched Sample.}\label{sec:deltas_measurement}
\subsection{High Energy Proton-Enriched Sample.}
As mentioned above, the $\Delta s$ sensitivity of the NCE sample comes down
to the possibility of distinguishing proton from neutron events.
In particular, the ratio of $\nu p\to \nu p$ to $\nu n\to \nu n$ is most sensitive to $\Delta s$ 
in addition to reducing the systematic error.
One should mention that even though FSI effects may be significant in the NCE cross section,
they become negligible when using such ratios~\cite{Meucci_FSI_in_the_ratio}

In MiniBooNE, NCE scattering from a neutron can only be detected when that neutron has a further strong interaction, usually with a proton.
At low energies, these cannot be distinguished from NCE scattering from a proton.
When scattering kinematics produces a proton above Cherenkov threshold, it is distinctive since the secondary interaction required to detect a neutron rarely produces an above-Cherenkov-threshold proton.
We thus select a special class of NCE protons, with only one proton in the final state whose energy is above Cherenkov threshold.
These \textit{single proton} events will be used for a $\Delta s$ measurement.

It should be noted that events with multiple nucleons in the final state (both
NCE neutrons and NCE-like background) have \textit{multiple protons} produced in the
reaction.
At kinetic energies above 350~MeV, single proton events produce on average
a higher Cherenkov light fraction than multiple proton events.
In addition, these two classes of events differ in the kinematics of the outgoing nucleon,
with single proton events being more forward-going.

\subsection{Ratio of $\nu p\to \nu p$ to $\nu N\to \nu N$.}
\begin{figure}
\includegraphics[bb = 0 0 565 385, width=\columnwidth]{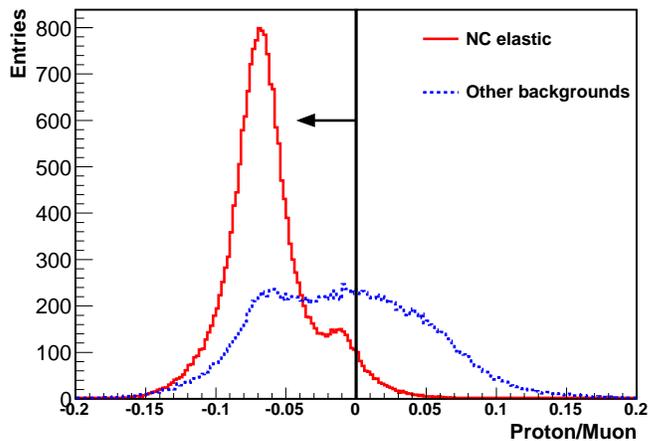}
\caption{The ``Proton/Muon'' cut based on the $\ln(\mathcal{L}_{\mu}/\mathcal{L}_{p})$ variable,
         which is the log-likelihood ratio of muon and proton event hypotheses.
         The left part of the cut is used for the analysis.
         The histograms are normalized to POT.
         } 
\label{fig:pr_other_proj}
\end{figure}

In order to measure  $\Delta s$, the ratio of $\nu p\to \nu p$ to $\nu N\to \nu N$ as a function of the reconstructed 
nucleon kinetic energy from 350~MeV to 800~MeV is used.
Additionally, by taking the ratio, several sources of systematic uncertainty are reduced.
The denominator of this ratio are events with NCE selection cuts 
described in Section \ref{sec:cuts}, but with the energy cut (5) replaced with
$350\mbox{ MeV}<T<800\mbox{ MeV}$,
and an additional ``Proton/Muon'' cut based on $\ln(\mathcal{L}_{\mu}/\mathcal{L}_{p})$
(the log-likelihood ratio between muon and proton hypotheses) in order to
reduce muon-like backgrounds that dominate the high visible energy region (Fig.~\ref{fig:pr_other_proj}).
After these cuts, the $\nu N\to \nu N$ data sample includes 24,004 events with the
following predicted channel fractions: 45\% NCE, 26\% NCE-like background,
3\% dirt background, and 25\% other backgrounds.

\begin{figure}
\includegraphics[bb = 0 0 565 385, width=\columnwidth]{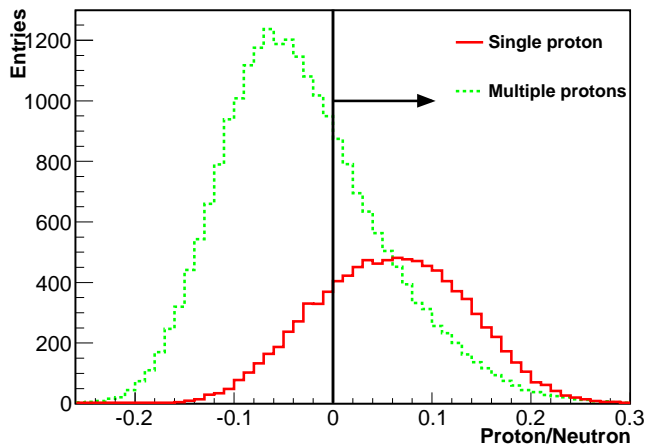}
\caption{The ``Proton/Neutron'' cut based on the fraction of prompt hits with
         the corrected time between 0~ns and 5~ns.
         The right part of the cut is used for the analysis.
         The histograms are normalized to POT.
         } 
\label{fig:pr_nn_proj}
\end{figure}

\begin{figure}
\includegraphics[bb = 0 0 565 385, width=\columnwidth]{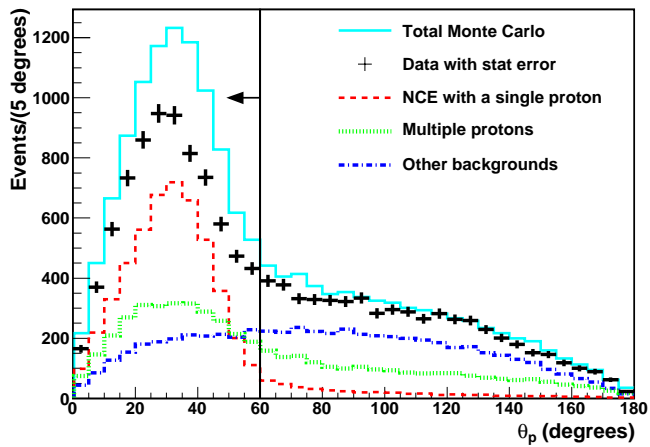}
\caption{The angle between the reconstructed nucleon direction and the incident
         beam direction for data and MC after the NCE, ``Proton/Muon'', and  ``Proton/Neutron'' cuts.
         All distributions are POT--normalized.
         Events with $\theta_p<60^{o}$ are used in the analysis.}
\label{fig:thetap}
\end{figure}

For the numerator of the ratio, two more cuts are applied in addition to the ones used for
the denominator.
The first of them, the ``Proton/Neutron'' cut, is a variable based on the
fraction of prompt light (as described in Section III.B, but with a corrected
time between 0~ns and 5~ns). This cut is to increase the single proton event fraction
in the sample. The described variable distribution and the value of the cut are shown in
Fig.~\ref{fig:pr_nn_proj} for both single proton and multiple proton events. 
As one can see, the cut reduces multiple proton events, which have less Cherenkov light than
single proton events. 
Finally, we apply a cut on the angle between the reconstructed nucleon direction and the incident beam direction.
As shown in Fig.~\ref{fig:thetap}, single proton events are mostly forward going; we thus require $\theta_p<60^o$.
The final $\nu p\to \nu p$ data sample includes 7,616 events with the following
predicted channel fractions:
55\% $\nu p\to \nu p$, 10\% $\nu n \to \nu n$, 14\% NCE-like background, 1\% dirt background, and
19\% other backgrounds.

\begin{figure}
\includegraphics[bb=0 0 565 385, width=\columnwidth]{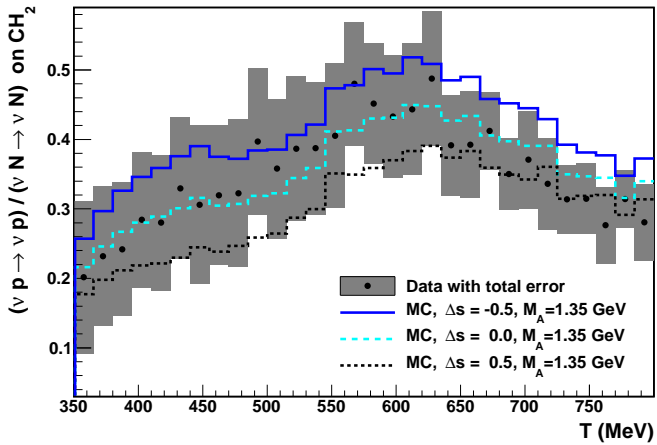}
\caption{The ratio of $\nu p\to \nu p /\nu N\to \nu N$ as a function of the
         reconstructed energy for data and MC with $\Delta s$ values as labeled.}
\label{fig:Data_deltas}
\end{figure}

The ratio of $\nu p\to \nu p$ to $\nu N\to \nu N$ events for data and MC for
different values of $\Delta s$ ($-0.5$, 0, and $+0.5$) is shown in
Fig.~\ref{fig:Data_deltas}, which illustrates the sensitivity of this ratio to
$\Delta s$.
The error bars for the data histogram are the diagonal elements of the full
error matrix.

\subsection{Measurement of $\Delta s$.}
The full error matrix is used for the $\chi^2$ tests of $\Delta s$ to determine the best fit and the confidence interval.
The $\chi^2$ surface also slightly depends on the value of $M_A$.
Assuming $M_A=1.35$~GeV, the fit to the MiniBooNE measured $\nu p\to\nu p$/$\nu N\to\nu N$ ratio yields:

\begin{equation}
\Delta s=0.08\pm0.26,
\label{eq:deltas_result}
\end{equation}
with $\chi^2_{min}/DOF=34.7/29$.
Using $M_A=1.23$~GeV yields $\Delta s=0.00\pm0.30$ with $\chi^2_{min}/DOF=34.5/29$.
The result is consistent with the BNL E734 measurement~\cite{BNL734}.

One needs to comment on the implications of Figures \ref{fig:erec}, \ref{fig:thetap}, and \ref{fig:Data_deltas} on the $\Delta s$ result.
From Fig.~\ref{fig:erec}, one can see that the MiniBooNE MC overpredicts the
total number of events passing the NCE selection cuts at high reconstructed
energies ($T>250\,\mbox{MeV}$) by as much as 40\%.
The NCE proton-enriched sample was obtained for events with
high reconstructed energies, $350<T<800$~MeV.
For these events, looking at the $\theta_p$ distribution in
Fig.~\ref{fig:thetap}, it seems that the entire disagreement between data
and MC at these energies comes from the forward-going events, $\nu p\to\nu p$.
Clearly there is a deficit of $\nu p \to \nu p$ in the data, which in principle implies
that this might be due to positive values of $\Delta s$.
However, Fig.~\ref{fig:Data_deltas} and the result in
Eq.\eqref{eq:deltas_result} shows that it is consistent with zero.
This may indicate that there is also a deficit of $\nu n \to \nu n$.

This measurement represents the first attempt at a $\Delta s$ determination using this ratio.
The systematic errors are quite large, mostly due to large uncertainties in the
optical model of the mineral oil.
MiniBooNE maintains a sensitivity to $\Delta s$ for proton energies above Cherenkov
threshold for protons, where the contribution from NCE-like background is significant.
In order to improve the sensitivity to $\Delta s$, future experiments
need to have good proton/neutron particle identification 
(possibly through neutron capture tagging)
and extend the cross-section measurement down to the $T<200$~MeV region, 
where the contribution from NCE-like background becomes 
negligible and where the extrapolation
of the axial form factor to $Q^2=0$ becomes less model dependent.

\section{Summary.}\label{sec:summary}
In summary, MiniBooNE has used a high-statistics sample of NCE interactions
to measure the NCE ($\nu N\to\nu N$) flux-averaged differential cross section, $d\sigma/dQ^2$ on CH$_2$.
Using MiniBooNE CCQE data, a measurement of NCE/CCQE cross-section ratio has also been performed.
Using POT-normalized distributions of the reconstructed energy for the
NCE sample, $\chi^2$ tests for several $M_A$ and $\kappa$ values have been
performed.
The MC with higher values of $M_A$ give a better $\chi^2$ than that with $M_A=1.02$~GeV.
The allowed region for the axial vector mass using just MiniBooNE NCE data was obtained, $M_A=1.39\pm0.11$~GeV.
It is in agreement with the shape normalized fits of $\nu_\mu$ CCQE scattering on neutrons bound in carbon, oxygen and iron 
as obtained by recent experiments~\cite{MB_CCQE_PRD,K2K_ccqe, K2K_ccqe2,MinosMA_Dorman_Thesis}.
For energies above Cherenkov threshold, a sample of NCE proton-enriched events
was obtained, which was used for the measurement of the $\nu p\to\nu p$ to $\nu N\to\nu N$ ratio, 
which in turn is sensitive to $\Delta s$.
A value of $\Delta s=0.08\pm0.26$ was extracted, in agreement with
the results from the BNL E734 experiment~\cite{BNL734}.



\bibliographystyle{apsrev}

\bibliography{ncel_prd}


\appendix
\section*{Appendix A: Dirt Background Measurement.}\label{sec:dirt_measurement}
%

The fact that dirt and in-tank events have different spatial distributions, such as reconstructed radius, $R$, and the reconstructed $Z$ coordinate (which is in the direction of the beam)
can be used to determine the dirt energy spectrum.
Dirt event vertices are generally reconstructed closer to the edge of the detector, than the in-tank events.
Also, dirt events are mostly reconstructed in the upstream part of the detector  with $Z<0$, whereas in-tank events have approximately a uniform distribution in this variable.
Additionally, we use the fact that dirt and in-tank events have a very different energy spectrum as seen in Fig.\ref{fig:erec}.

The dirt energy spectrum is measured by fitting the MC in-tank and dirt
templates to the data in $Z$, $R$, and energy distributions for the dirt-enriched event samples.
The $R$ and $Z$ distributions are obviously correlated.
However, the samples used for the $Z$ and $R$ fits of the dirt have a large
fraction of events that are present in one sample and not in the other.

To measure the dirt background in the NCE event sample, three additional dirt-enriched samples of events are used.
For each of the variables (reconstructed $Z$, $R$ and energy) different
samples are used, based on cutting on the other of these variables;
for example, if measuring dirt events from the $Z$ distribution, one would have an additional cut on $R$.

The samples that are selected for these fits are defined in Table~\ref{tab:dirtsamples}.
The following precuts are the same for each event sample (cuts from (1) through
(6) described in Section \ref{sec:cuts}):
$$
\begin{array}{ccl}
\mbox{Precuts} &=& 1\mbox{ Subevent} + \mbox{VHits}<6 + \mbox{THits}>24 \\
               &+& 4400\mbox{ ns}<\mbox{Time}<6500\mbox{ ns} + T<650\mbox{ MeV}\\
	       &+& \ln(\mathcal{L}_{e}/\mathcal{L}_{p})<0.42 .
\end{array}
$$
These are the same as the NCE analysis cuts with the removal of the radial cut.
The dirt fraction in the dirt-enhanced samples is increased significantly, by a factor of 2--3
over the unenhanced sample.

\begin{table*}[hbt!]
\begin{center}
  \begin{tabular}{lllr}
    \hline
    \hline
    Sample  name & Purpose of the sample             & Cuts : $Precuts +$                   & Dirt fraction (\%)   \\
    \hline
    NCE       & NCE sample (dirt-reduced)             & $R_{fiducial} (T)$ (cut (7) in Section~\ref{sec:cuts})                          & 13.4  \\
    Dirt--Z   & Fit dirt from Z (dirt-enhanced)           & $3.8\mbox{ m}<R<5.2\mbox{ m}$                                          & 27.8  \\
    Dirt--R   & Fit dirt from R (dirt-enhanced)          & $Z<0\mbox{ m}$                                                                       & 34.3  \\
    Dirt--E   & Fit dirt from energy (dirt-enhanced) & $3.8\mbox{ m}<R<5.2\mbox{ m}$ and $Z<0\mbox{ m}$  & 37.6 \\
    \hline
    \hline
  \end{tabular}
\end{center}
\caption{Event sample cuts, their respective purposes, and dirt events fractions.
         The dirt fractions are calculated from the initial MC simulation
         (before dirt fits), NCE is the signal sample (cuts from 1 to 7),
         whereas the other three samples are the dirt-enriched samples for use
         in the dirt fits.}
\label{tab:dirtsamples}
\end{table*}

\subsubsection{Dirt Rate Measurement from the Reconstructed Z Distribution.}
To measure the dirt background from the $Z$ distribution, the ``Dirt--Z''
event sample from Table~\ref{tab:dirtsamples} is used.
The shapes (templates) of the $Z$ distribution for the in-tank and dirt events
in the MC are used to fit the shape of the $Z$ distribution for the data.
These fits are done in bins of reconstructed energy $T$, so that in the end
one obtains the measured dirt background energy spectrum.
From the fits the correction function 
$$
\mbox{Corr}^{Dirt}_{i} = 
\left(
Data_i/MC_i
\right)^{Dirt}
$$ 
that should be
applied to the MC spectrum of the dirt events is extracted, which is the ratio
of the measured to predicted number of dirt events in the $i$-th bin.

\begin{figure*}
\includegraphics[height=5.2in, width=7.0in]{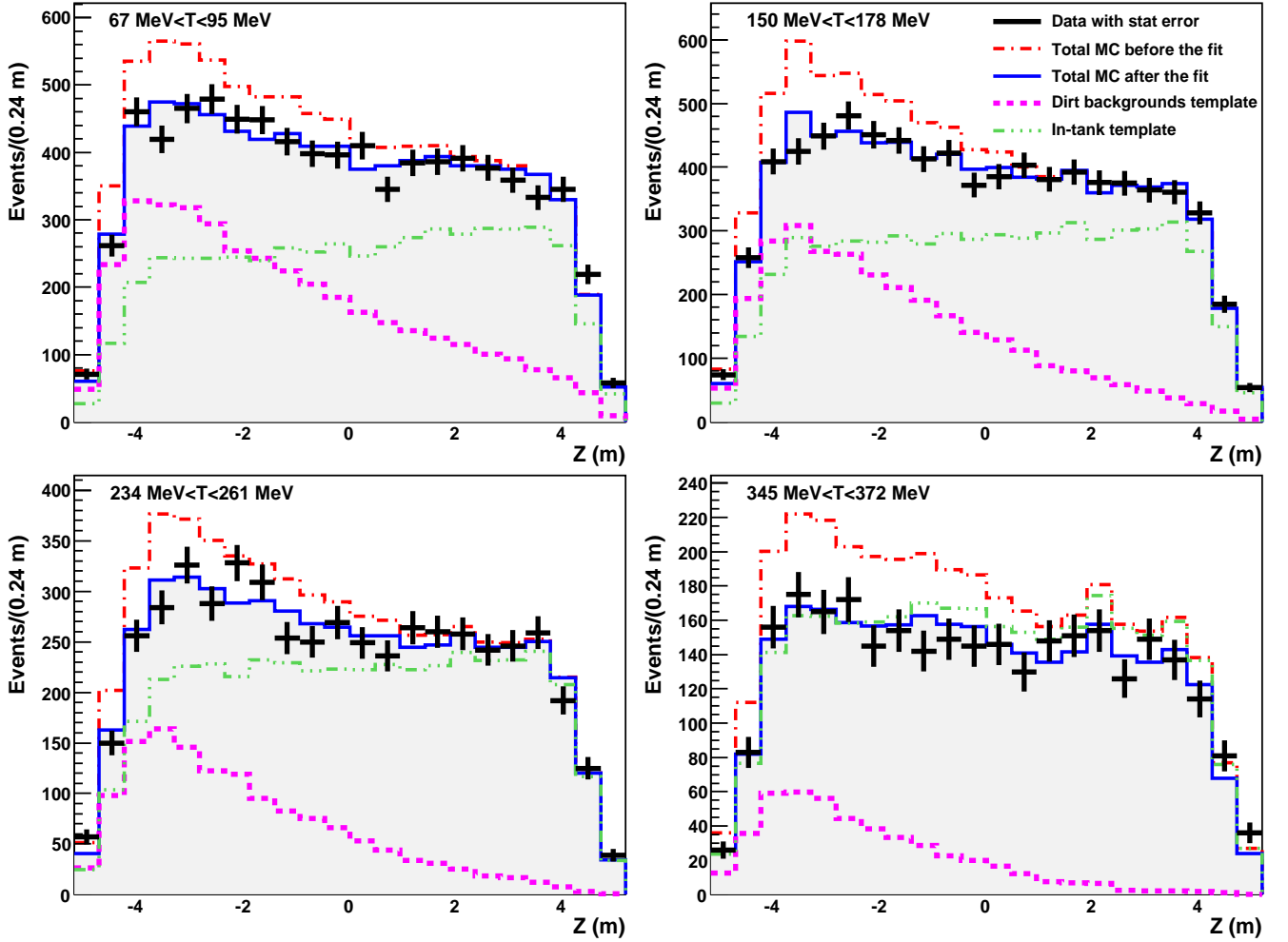}
\caption{Fits to the data using MC templates for the in-tank and dirt events in Z  for different reconstructed energy bins.
	The ``total MC before the fit" is a sum of the in-tank and dirt templates, which are absolutely (POT) normalized.
	}
\label{fig:DirtZ_fit}
\end{figure*}

An example of the fit is shown in Fig.~\ref{fig:DirtZ_fit} for several
intervals of the reconstructed energy.
The agreement between the data and MC is much better after the fit.

\subsubsection{Dirt Rate Measurement from the Reconstructed R Distribution.}
This procedure is essentially the same as for the dirt measurement using the
$Z$ distribution, but instead, the ``Dirt--R'' sample from
Table~\ref{tab:dirtsamples} is used and the fitting is done for the $R$
variable.
Again, from the fits, the correction function for the dirt energy spectrum
$\mbox{Corr}^{Dirt}$ is extracted.

\subsubsection{Dirt Rate Measurement from the Reconstructed Energy Distribution.}
For this method, two event samples are used, the signal (NCE) and the dirt-enriched sample with the ``Dirt--E'' cuts from Table~\ref{tab:dirtsamples}.
For both samples, we look at both the MiniBooNE data and the MC prediction for
NCE, dirt and in-tank backgrounds.

Assuming that the fractions of signal and dirt events in both samples are stable relative to MC variations, one can measure the spectrum of dirt events in the NCE
sample from the data distribution for both of these samples.
We define the following histograms:
$$
\begin{array}{lcl}
\nu &-& \mbox{ reconstructed energy spectrum for data}\\
B   &-& \mbox{ reconstructed energy spectrum for MC in-tank}\\
    & & \mbox{ backgrounds}\\
S   &-& \mbox{ reconstructed energy spectrum for MC NCE}\\
D   &-& \mbox{ reconstructed energy spectrum for MC dirt},
\end{array} 
$$
which have upper indices describing the event sample, namely
$$
\begin{array}{lcl}
s &-& \mbox{ NCE event sample}\\
d &-& \mbox{ ``Dirt--E'' event sample}.
\end{array} 
$$

The new spectra for the dirt and signal events can be determined from fitting
the data in both NCE and ``Dirt--E'' samples.
In terms of the definitions that we have introduced, the condition that these
spectra coincide in both event samples can be written as:
\begin{equation}
\label{eq:dirtenergy_soe}
\begin{cases}
     B_i^s + S_i^s + D_i^s & = \,\nu_i^s \\
     B_i^d + S_i^d + D_i^d & = \,\nu_i^d.
\end{cases}
\nonumber
\end{equation}
For each reconstructed energy bin $i$ there are 6 unknowns on the left hand
side and 2 knowns on the right hand side (the data in both NCE and ``Dirt--E''
samples).
Assuming a reliable in-tank background prediction, one can fix $B_i^s$ and $B_i^d$.
Then, we introduce the fractions of signal and dirt events in the two samples:
\begin{equation}
\displaystyle f_i = \frac{\displaystyle D^d_i}{\displaystyle D^s_i}\qquad \mbox{and}\qquad g_i  = \frac{\displaystyle S^d_i}{\displaystyle S^s_i}.
\nonumber
\end{equation}
Because these variables are ratios, they are relatively stable to MC variations and independent of the dirt and NCE events energy spectra.
The functions $f$ and $g$ are determined from the MC.

Herewith, one can express $D_i^s$ (the dirt energy spectrum in the NCE sample)
in terms of the above definitions as:
\begin{equation}
D_i^s=\frac{\displaystyle g_i\left(\nu^s_i-B^s_i\right)-\left(\nu^d_i-B^d_i\right)}{\displaystyle g_i-f_i},
\nonumber
\end{equation}
which is the measured spectrum of dirt events in the NCE event sample.

\begin{figure}[t!]
\includegraphics[height=2.6in, width=3.5in]{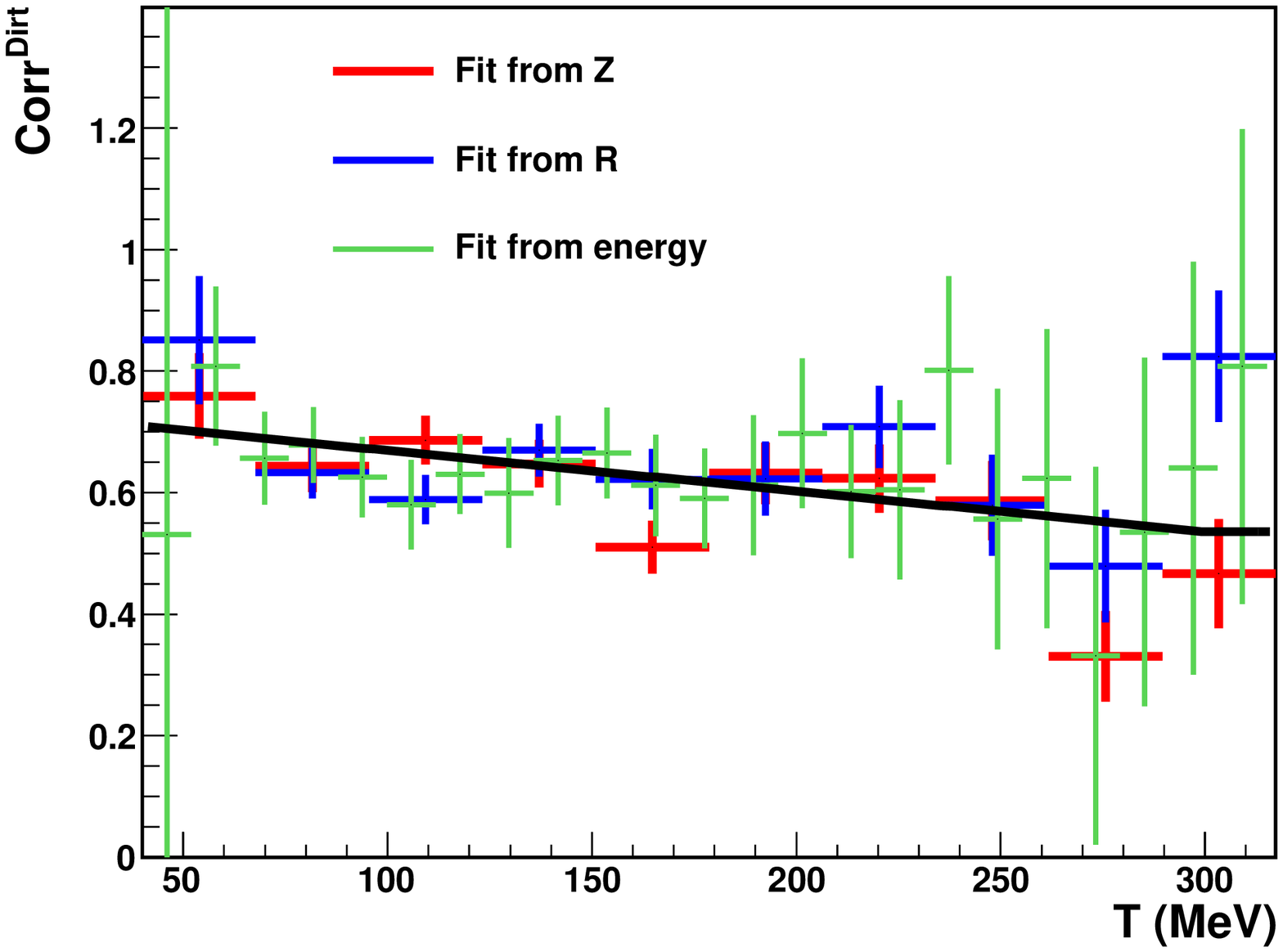}
\caption{Combined dirt energy correction fit from $Z$, $R$ and energy distributions. Errors for $Z$ and $R$ are statistical from the minimization fitting these distributions.
The error for the energy distribution is the largest systematic uncertainty resulting from the optical model of the mineral oil.
}
\label{fig:DirtCorrection}
\end{figure}

Finally, using all three methods, the combined dirt energy spectrum correction
function fit is performed.
The chosen form of the fit function is linear below 300~MeV and a constant
above, as shown in Fig.~\ref{fig:DirtCorrection}.
All of these measurements agree with each other within 10\%, lending confidence to
our overall ability to constrain the dirt background in the analysis.

The new, measured reconstructed energy spectrum for the dirt events in the NCE
sample is calculated as a bin-by-bin correction of the initial MC dirt energy
spectrum multiplied by the measured correction function.
The measured number of dirt events is determined to be $\sim 30\%$ lower than
the original MC prediction.

\appendix
\section*{Appendix B: MiniBooNE Neutral--Current Elastic Cross-Section Description.}\label{sec:xs_description}

\subsection*{B.1 Phenomenology.}\label{sec:phenom}

While MiniBooNE uses relativistic Fermi gas model of Smith and Moniz to describe NCE scattering,
it is illustrative to write down the expression for the NCE cross-section in case of a free nucleon target.

The neutrino-nucleon NCE differential cross-section on free nucleons can be written as~\cite{Garvey1}:
\begin{equation}\label{eq:ncel_xs_theory_formula}
\frac{\displaystyle d\sigma}{\displaystyle dQ^2}=\DP\frac{\DP G_F^2Q^2}{\DP 2\pi E_\nu^2}\left(A(Q^2)\pm B(Q^2)W+C(Q^2)W^2\right),
\end{equation}
where the '$+$' sign corresponds to neutrinos and the '$-$' sign to anti-neutrinos, 
$W={4E_\nu}/{M_N}-{Q^2}/{M_N^2}$, and $A(Q^2)$, $B(Q^2)$ and $C(Q^2)$ are form factors defined as:
\begin{widetext}
\begin{equation}\label{eq:ncel_xs_theory_formula_FF}\nonumber
\begin{array}{lcl}
  A(Q^2)        & = & \displaystyle\frac14 \left[(F_A^Z)^2(1+\tau)-((F_1^Z)^2-\tau (F_2^Z)^2)(1-\tau)+4\tau F_1^ZF_2^Z  \right]\makespace,\\
  B(Q^2)        & = & \displaystyle\frac14 F_A^Z(F_1^Z+F_2^Z)\makespace, \\
  C(Q^2)        & = & \displaystyle\frac{\displaystyle M_N^2}{\displaystyle 16Q^2} \left[ (F_A^Z)^2+(F_1^Z)^2+\tau(F_2^Z)^2 \right].
  \end{array}
\end{equation}
Here $F_1^Z$, $F_2^Z$, and $F_A^Z$ are nucleon Dirac, Pauli and axial weak neutral-current 
form factors, respectively, which in general are real dimensionless functions of $Q^2$, and $\tau={Q^2}/{4M_N^2}$.
Each of the nucleon for factors is different for proton and neutron targets.
At low $Q^2$ the $C(Q^2)$ term in Eq.\eqref{eq:ncel_xs_theory_formula} dominates (see Ref.\cite{Perevalov_thesis}).
Thus, the NCE cross-section has a significant contribution from axial vector currents.

Under the conserved vector current~\cite{Pachos_text}, one can express the weak form factors through their electromagnetic equivalents:

\begin{equation}\label{eq:FF_Z2}
F_i^{Z}=\left(\displaystyle\frac12-\sin^2\theta_W\right)\left[F_i^{EM,p}-F_i^{EM,n}\right]\tau_3  -
\sin^2\theta_W\left[F_i^{EM,p}+F_i^{EM,n}\right] - \displaystyle\frac12 F_i^s , \qquad i=1,2,
\end{equation}
\end{widetext}
where $\theta_W$ is the weak mixing angle, $\tau_3$ is a factor +1 for protons and -1 for neutrons, 
$F_i^s$ are the isoscalar form factors discussed later, and the Dirac and Pauli electromagnetic form factors are:

\begin{equation}\label{eq:FF_EM}\nonumber
\begin{array}{lcl}
F_1^{EM}(Q^2) &=& \displaystyle{\frac{ \displaystyle G_E(Q^2) +\frac{\displaystyle Q^2}{\displaystyle 4M_N^2}G_M(Q^2)}{\displaystyle 1+\frac{\displaystyle Q^2}{\displaystyle 4M_N^2}}},\makespace\\
F_2^{EM}(Q^2) &=& \displaystyle{\frac{ \displaystyle G_M(Q^2)-G_E(Q^2)}{\displaystyle 1+\frac{\displaystyle Q^2}{\displaystyle 4M_N^2}}},
\end{array}
\end{equation}
where $G_E$ and $G_M$ are Sachs form factors~\cite{Sachs_FF}.
Instead of the dipole approximation the BBA-2003 form of the Sachs form factors~\cite{BBA03} is used
in this analysis, which better describes the electron-proton scattering data.


The axial weak form factor by definition can be expressed through its isovector and isoscalar parts:

\begin{equation}\label{eq:FF_AZ}
F_A^Z=\displaystyle\frac{\DP \tau_3}{2}F_A-\displaystyle\frac{1}{2}F_A^s.
\end{equation}
The isovector axial form factor can be measured via weak charged-current.
Usually it is assumed to have a dipole form:

\begin{equation}\label{eq:FA}
F_A(Q^2)=\displaystyle\frac{\displaystyle g_A}{\left(\displaystyle 1+\frac{\displaystyle Q^2}{\displaystyle M_A^2}\right)^2},
\end{equation}
where $g_A=F_A(0)=1.2671$~\cite{PDG} is measured precisely from neutron beta decay.



The isoscalar form factors $F_1^s$ and $F_2^s$ in Eq.\eqref{eq:FF_Z2} are usually thought to be due to contributions from 
strange quarks to the electric charge and to the magnetic moment of the nucleon, whereas
$F_A^s$ in Eq.\eqref{eq:FF_AZ} is their contribution to the nucleon spin.
$F_A^s$ value at $Q^2=0$ is called $\Delta s$.
$F_1^s$ and $F_2^s$ can be extracted from parity violating electron scattering experiments.
Recent results from the HAPPEX experiment~\cite{HappexII} show that these electric and
magnetic strange form factors are consistent with zero.
The exact expression for the axial isoscalar form factor is unknown,
but in analogy to the isovector axial form factor it is usually represented in the dipole form 
with the same value of the axial vector mass to minimize the number of free parameters in the model~\cite{Garvey_Louis}:

\begin{equation}\label{eq:strange_FF}
\begin{array}{lcl}
F_A^s(Q^2)&=&\displaystyle{\frac{\displaystyle\Delta s}{\left(\displaystyle 1+\frac{\displaystyle Q^2}{\displaystyle M_A^2}\right)^2}}.
\end{array}
\end{equation}

\subsection*{B.2 MiniBooNE Neutral--Current Elastic Cross-Section Discussion.}
The MiniBooNE NCE scattering sample consists of three different processes: scattering on
free protons in hydrogen, bound protons in carbon, and bound neutrons in carbon.
Because several final state nucleons may be produced, 
we define the interaction in carbon using most energetic final state nucleon.
This means, for example, that it is possible for an event to be tagged as a NCE neutron, 
because it has a neutron as the most energetic final state nucleon, 
even though the original neutrino interaction was on a proton.
According to NUANCE, the probability of this misidentification grows almost linearly 
from about $8\%$ at $Q^2_{QE} = 0.1\mbox{GeV}^2$ to $16\%$ at $Q^2_{QE}=1.6\mbox{ GeV}^2$.

\begin{figure}[t!]
\includegraphics[bb = 0 0 560 385, width=0.5\textwidth]
                {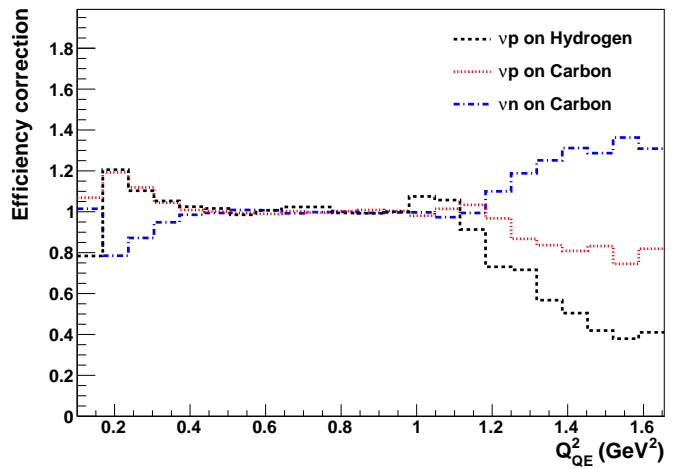}
\caption{Efficiency corrections $C_{\nu p,H}$, $C_{\nu p,C}$, and $C_{\nu n,C}$
         versus $Q^2_{QE}$ for different NCE processes (as labeled).}
\label{fig:Correction_coeff_FlatFiducial}
\end{figure}

The result shown in Fig.~\ref{fig:xs} is the flux-averaged NCE
differential cross-section on CH$_2$, averaged over these processes.
Herewith, the $\nu N\to \nu N$ cross-section is expressed as:
\begin{eqnarray}\label{eq:NCE_xs_expression}
   \frac{\displaystyle d\sigma_{\nu N\to \nu N}}{\displaystyle dQ^2} &=&
   \frac{1}{7}\,C_{\nu p,H}(Q^2_{QE})\,\frac{\displaystyle d\sigma_{\nu p\to\nu p,H}}{\displaystyle dQ^2}
   \nonumber\\
                                         &+&
   \frac{3}{7}\,C_{\nu p,C}(Q^2_{QE})\,\frac{\displaystyle d\sigma_{\nu p\to\nu p,C}}{\displaystyle dQ^2}\\
                                         &+&
   \frac{3}{7}\,C_{\nu n,C}(Q^2_{QE})\,\frac{\displaystyle d\sigma_{\nu n\to\nu n,C}}{\displaystyle dQ^2},
   \nonumber
\end{eqnarray}
where $d\sigma_{\nu p\to \nu p,H}/dQ^2$ is the NCE cross-section on free
protons (per free proton), $d\sigma_{\nu p\to \nu p,C}/dQ^2$ is the NCE cross-section on bound
protons (per bound proton), and $d\sigma_{\nu n\to \nu n,C}/dQ^2$ is the NCE
cross-section on bound neutrons (per bound neutron).
The efficiency correction functions $C_{\nu p,H}$, $C_{\nu p,C}$, and
$C_{\nu n,C}$ result from different selection efficiencies for each type of NCE
scattering process, and are estimated from the MC as functions of $Q^2_{QE}$ -- as
shown in Fig.~\ref{fig:Correction_coeff_FlatFiducial}.
They are defined as the ratios of the efficiency for a particular type of NCE
event to the average efficiency for all NCE events in bins of $Q^2_{QE}$.

As one can see from Fig.~\ref{fig:Correction_coeff_FlatFiducial},
the efficiency correction functions are equal to unity in the region of $Q^2_{QE}$ from 0.4 to $1.2\mbox{ GeV}^2$. 
However, at higher $Q^2_{QE}$, NCE neutrons have higher efficiency, thus having a higher probability
than NCE protons to pass the $T<650$~MeV cut used in the selection.
Similarly, at lower $Q^2_{QE}$, NCE neutrons have lower probability than NCE protons to pass
a $THits>24$ cut.

To calculate a cross-section which is to be compared to the MiniBooNE results
(as in Fig.~\ref{fig:xs}), one needs to apply these efficiency corrections to
each predicted distribution.
However, note that in the bulk of the measured region,
$0.1<Q^2_{QE}<1.0\,\mbox{GeV}^2$, they are all roughly equivalent to unity.

Finally, we present tables with the NCE/CCQE differential cross-section ratios in Table~\ref{tab:nce_ccqe} 
and the NCE differential cross-section, NCE-like background and correction coefficients in Table~\ref{tab:sigma_nce}.
There is also an alternative method of reporting the results of this analysis, 
namely in terms of the reconstructed nucleon kinetic energy.
The tables for the latter can be found in \cite{Perevalov_thesis}~and~\cite{MB_opendata_nce}.
In order to make use of these results, one would have to follow 
the instructions described in Appendix~B of Ref.~\cite{Perevalov_thesis}.

\begin{table}
\begin{center}
\small\addtolength{\tabcolsep}{-0.1pt}
\resizebox{0.5\textwidth}{!} {
\begin{tabular}{c | c c}
\hline
\hline
\raisebox{-0.5ex}{$Q^2(\mbox{GeV}^2)$}\raisebox{0.2ex}{$\backslash$}\raisebox{0.5ex}{Distribution}& $\frac{\DP\sigma_{NCE}}{\DP\sigma_{CCQE}}$ & $\frac{\DP\sigma_{NCE-like}}{\DP\sigma_{CCQE-like}}$  \\ 
\hline
0.100--0.150 & $(2.019 \pm 0.261)\times 10^{-1}$ & $(1.682 \pm 0.199)\times 10^{-1}$  \\  
0.150--0.200 & $(1.839 \pm 0.225)\times 10^{-1}$ & $(1.582 \pm 0.181)\times 10^{-1}$  \\  
0.200--0.250 & $(1.769 \pm 0.199)\times 10^{-1}$ & $(1.554 \pm 0.165)\times 10^{-1}$  \\  
0.250--0.300 & $(1.696 \pm 0.173)\times 10^{-1}$ & $(1.526 \pm 0.147)\times 10^{-1}$  \\  
0.300--0.350 & $(1.619 \pm 0.167)\times 10^{-1}$ & $(1.502 \pm 0.145)\times 10^{-1}$  \\  
0.350--0.400 & $(1.620 \pm 0.192)\times 10^{-1}$ & $(1.561 \pm 0.172)\times 10^{-1}$  \\  
0.400--0.450 & $(1.594 \pm 0.231)\times 10^{-1}$ & $(1.601 \pm 0.210)\times 10^{-1}$  \\  
0.450--0.500 & $(1.602 \pm 0.264)\times 10^{-1}$ & $(1.680 \pm 0.244)\times 10^{-1}$  \\  
0.500--0.600 & $(1.584 \pm 0.313)\times 10^{-1}$ & $(1.767 \pm 0.295)\times 10^{-1}$  \\  
0.600--0.700 & $(1.532 \pm 0.416)\times 10^{-1}$ & $(1.855 \pm 0.397)\times 10^{-1}$  \\  
0.700--0.800 & $(1.498 \pm 0.572)\times 10^{-1}$ & $(1.966 \pm 0.554)\times 10^{-1}$  \\  
0.800--1.000 & $(1.421 \pm 0.755)\times 10^{-1}$ & $(2.028 \pm 0.737)\times 10^{-1}$  \\  
1.000--1.200 & $(1.408 \pm 0.712)\times 10^{-1}$ & $(2.102 \pm 0.720)\times 10^{-1}$  \\  
1.200--1.500 & $(1.226 \pm 0.643)\times 10^{-1}$ & $(1.897 \pm 0.666)\times 10^{-1}$  \\  
1.500--2.000 & $(8.948 \pm 5.248)\times 10^{-2}$ & $(1.431 \pm 0.736)\times 10^{-1}$  \\  
\hline
\hline
\end{tabular}}
\end{center}
\caption{
NCE/CCQE and NCE-like/CCQE-like differential cross-section ratios as a function of $Q^2_{QE} = 2m_N\sum_i T_i$.}
\label{tab:nce_ccqe}
\end{table}
\begin{table*}
\begin{center}
\small\addtolength{\tabcolsep}{-0.1pt}
\begin{tabular}{c | c c c c c}
\hline
\hline
\raisebox{-1.0ex}{$Q^2(\mbox{GeV}^2)$}\raisebox{-0.25ex}{$\backslash$}\raisebox{0.5ex}{Distribution}& NCE cross-section, & NCE-like background, & $C_{\nu p,H}$ & $C_{\nu p,C}$  & $C_{\nu n,C}$ \\ 
& cm$^2$/GeV$^2$ & cm$^2$/GeV$^2$ &   & \\ \hline
0.101--0.169 & $(3.361 \pm 0.360)\times 10^{-39}$ & $4.875\times 10^{-41}$ &0.784 &1.068 &1.014 \\  
0.169--0.236 & $(2.951 \pm 0.394)\times 10^{-39}$ & $4.623\times 10^{-41}$ &1.206 &1.192 &0.785 \\  
0.236--0.304 & $(2.494 \pm 0.429)\times 10^{-39}$ & $5.879\times 10^{-41}$ &1.102 &1.119 &0.871 \\  
0.304--0.372 & $(2.089 \pm 0.340)\times 10^{-39}$ & $8.425\times 10^{-41}$ &1.053 &1.043 &0.949 \\  
0.372--0.439 & $(1.744 \pm 0.243)\times 10^{-39}$ & $1.235\times 10^{-40}$ &1.024 &1.009 &0.985 \\  
0.439--0.507 & $(1.432 \pm 0.246)\times 10^{-39}$ & $1.647\times 10^{-40}$ &1.016 &1.002 &0.994 \\  
0.507--0.574 & $(1.168 \pm 0.260)\times 10^{-39}$ & $1.964\times 10^{-40}$ &0.986 &0.994 &1.009 \\  
0.574--0.642 & $(9.435 \pm 2.400)\times 10^{-40}$ & $2.155\times 10^{-40}$ &1.007 &0.989 &1.008 \\  
0.642--0.709 & $(7.534 \pm 2.205)\times 10^{-40}$ & $2.222\times 10^{-40}$ &1.023 &1.002 &0.992 \\  
0.709--0.777 & $(6.015 \pm 2.194)\times 10^{-40}$ & $2.215\times 10^{-40}$ &1.023 &0.995 &0.998 \\  
0.777--0.844 & $(4.832 \pm 2.320)\times 10^{-40}$ & $2.075\times 10^{-40}$ &0.994 &1.003 &0.999 \\  
0.844--0.912 & $(3.854 \pm 2.331)\times 10^{-40}$ & $1.890\times 10^{-40}$ &0.993 &1.009 &0.994 \\  
0.912--0.980 & $(3.209 \pm 2.330)\times 10^{-40}$ & $1.756\times 10^{-40}$ &0.999 &1.004 &0.997 \\  
0.980--1.047 & $(2.649 \pm 2.117)\times 10^{-40}$ & $1.521\times 10^{-40}$ &1.074 &0.980 &0.997 \\  
1.047--1.115 & $(2.226 \pm 1.818)\times 10^{-40}$ & $1.265\times 10^{-40}$ &1.056 &1.015 &0.973 \\  
1.115--1.182 & $(1.935 \pm 1.295)\times 10^{-40}$ & $1.081\times 10^{-40}$ &0.913 &1.034 &0.994 \\  
1.182--1.250 & $(1.598 \pm 0.939)\times 10^{-40}$ & $9.324\times 10^{-41}$ &0.731 &0.967 &1.099 \\  
1.250--1.317 & $(1.329 \pm 0.769)\times 10^{-40}$ & $8.079\times 10^{-41}$ &0.716 &0.867 &1.187 \\  
1.317--1.385 & $(1.111 \pm 0.689)\times 10^{-40}$ & $6.983\times 10^{-41}$ &0.567 &0.836 &1.251 \\  
1.385--1.452 & $(9.259 \pm 6.254)\times 10^{-41}$ & $5.958\times 10^{-41}$ &0.504 &0.808 &1.312 \\  
1.452--1.520 & $(7.975 \pm 5.373)\times 10^{-41}$ & $5.229\times 10^{-41}$ &0.419 &0.832 &1.286 \\  
1.520--1.588 & $(6.618 \pm 4.524)\times 10^{-41}$ & $4.342\times 10^{-41}$ &0.378 &0.746 &1.363 \\  
1.588--1.655 & $(5.799 \pm 3.921)\times 10^{-41}$ & $3.839\times 10^{-41}$ &0.410 &0.818 &1.309 \\  
\hline
\hline
\end{tabular}
\end{center}
\caption{
MiniBooNE measured NCE differential cross-section, predicted NCE-like background, and predicted correction coefficients for the three different NCE scattering contributions as a function of $Q^2_{QE}=2m_N\sum_i T_i$.}
\label{tab:sigma_nce}
\end{table*}

\IfFileExists{\jobname.bbl}{}
 {\typeout{}
  \typeout{******************************************}
  \typeout{** Please run "bibtex \jobname" to obtain}
  \typeout{** the bibliography and then re-run LaTeX}
  \typeout{** twice to fix the references!}
  \typeout{******************************************}
  \typeout{}
 }


\end{document}